# Fusion ignition via a magnetically-assisted fast ignition approach

**W.-M. Wang[1,3]*, P. Gibbon[2,4]†, Z.-M. Sheng[3,5,6]‡, Y. T. Li[1,3], and J. Zhang[3,6]**

[1]Beijing National Laboratory for Condensed Matter Physics, Institute of Physics, CAS, Beijing 100190, China

[2]Forschungszentrum Jülich GmbH, Institute for Advanced Simulation, Jülich Supercomputing Centre, D-52425 Jülich, Germany

[3]Collaborative Innovation Center of IFSA, Shanghai Jiao Tong University, Shanghai 200240, China

[4]Centre for Mathematical Plasma Astrophysics, Katholieke Universiteit Leuven, 3000 Leuven, Belgium

[5]SUPA, Department of Physics, University of Strathclyde, Glasgow G4 0NG, United Kingdom

[6]Key Laboratory for Laser Plasmas (MoE) and Department of Physics and Astronomy, Shanghai Jiao Tong University, Shanghai 200240, China

---

* E-mail: weiminwang1@iphy.ac.cn
† E-mail: p.gibbon@fz-juelich.de
‡ E-mail: zhengming.sheng@strath.ac.uk

**Significant progress has been made towards laser-driven fusion ignition via different schemes, including direct and indirect central ignition, fast ignition, shock ignition, and impact ignition schemes. However, to reach ignition conditions, there are still various technical and physical challenges to be solved for all these schemes. Here, our multi-dimensional integrated simulation shows that the fast-ignition conditions could be achieved when two 2.8 petawatt heating laser pulses counter-propagate along a 3.5 kilotesla external magnetic field. Within a period of 5 picoseconds, the laser pulses heat a nuclear fuel to reach the ignition conditions. Furthermore, we present the parameter windows of lasers and magnetic fields required for ignition for experimental test.**

To achieve controlled nuclear fusion energy, the central ignition concept of inertial confinement fusion (ICF) was proposed in 1970s [1-4]. With development of high power laser technology, in particular, with the operation of the National Ignition Facility in the United States, significant progress has been made in ICF [5]. However, more efforts are demanded to reach the ignition conditions. As another promising approach, the fast ignition (FI) concept was proposed in 1994 [6]. Different from the central ignition of ICF, it utilizes strong beams of MeV electrons generated by petawatt (PW) heating laser pulses [7] to heat the core region of a pre-compressed target after hundreds of micrometers transport. Extensive studies have been devoted to this scheme in the last two decades [8,9]. It was found that the large divergence of the generated electrons [8,10,11] significantly reduces the target heating efficiency. To mitigate this difficulty, a cone-inserted target was adopted [12], which has been very successful under certain conditions.

But different experiments showed widely variable heating efficiencies [12-15] under different conditions, e.g., different prepulse levels of the heating lasers, casting doubt on the efficacy of the cone-inserted scheme.

On the other hand, some recent technical developments allow one to examine the FI concept in completely new ways. Firstly, a few laser systems of PW and tens of kilojoules (kJ) are available recently or under construction worldwide [16], e.g., the LFEX system in Japan can deliver laser pulses of 2 PW and 10 kJ [17], with plans to be upgraded to 10 PW. At the same time, it was reported that ns-pulsed magnetic fields of 1.5 kilotesla (kT) can be produced by nanosecond-laser-driven capacitor coil experiments [18]. The generation of even higher magnetic fields is being pursued in a few laboratories. Moreover, about 8 kT magnetic fields have been generated during the compression of ICF targets initially imposed with moderate seed magnetic fields [19,20]. The combination of the two new technological capabilities in lasers and magnetic fields adds a new dimension to the FI concept. We have recently shown by integrated simulations that a 2 kT magnetic field imposed across a cone-free target can guide the transport of the divergent electrons generated by PW laser pulses and lead to significantly enhanced heating [21] to the target core even without a cone inserted into the target. Such a spherically symmetric target neither suffers from asymmetry in target compression [22], nor from possible preplasma side-effects appearing for a cone-inserted target [13-15]. However, whether and how the magnetically assisted FI scheme can realize ignition are still unclear and demand further investigation — this is the objective of the present work.

To achieve ignition via the FI scheme, the compressed target core region needs to be heated by the laser-generated fast electrons to a temperature high enough, e.g., 5 keV for the core with an areal density of 0.8 g cm$^{-2}$ [23], within the time limit of the compression disassembly, i.e., usually below 10 ps [6,12]. Hence, laser pulses with sufficient high powers are needed. This, however, may result in the generation of fast electrons with too high energies [24-26], such that they are not able to efficiently deposit their energy in the core. Here, we show by the full scale of integrated particle-in-cell (PIC) simulations that 0.35 μm (3ω) or 0.53 μm (2ω) heating laser pulses need to be taken to reduce the electron energy and enhance the laser-to-core coupling, and can realize fusion ignition via the magnetically assisted scheme. While 1.05 μm fundamental heating laser pulses are taken, it is hard to realize ignition. Our integrated simulation shows that two counter-propagating, 2.8 PW laser pulses with the total energy of 28 kJ convert 15% their energy to a target core. The latter is heated up to above 5keV within 5 ps, reaching the ignition condition [23]. Here, a 3.5 kT magnetic field is imposed across the cone-free target with the core of an density of 400 g cm$^{-3}$ and an areal density of 0.8 g cm$^{-2}$. Moreover, a parameter window for fusion ignition is presented under different laser powers and magnetic field strengths.

**Fusion ignition predicted by integrated simulation**

To exactly calculate the target heating, the electron generation should be computed in real time and then coupled with the electrons transport self-consistently. Note that the energy spectrum of the generated electrons changes with time due to the temporally-varying laser-plasma interaction, with which the target heating should be different from the case just using final energy spectrum after the laser-plasma interaction, an approximation that has been adopted by most simulation studies reported so far. Here, we take the ``two-system'' integrated PIC simulation approach [27] (see Methods), which self-consistently includes the whole FI heating processes from electron generation by laser pulses to fast electron transport and heating to a realistic pre-compressed target. The energy coupling from the heating lasers to the target core can also be directly obtained.

Considering the target design optimized for the FI scheme [28], we take a spherically symmetric deuterium-tritium (1:1 in number) target with an uniform density 400 g cm$^{-3}$ (or $9.5 \times 10^4$ $n_c$ with $n_c =10^{21}$ cm$^{-3}$) within a sphere of radius 20 μm (the core), and the surrounding density decreasing exponentially with a scalelength 9 μm along the radial direction from the density of 133.3 g cm$^{-3}$ at the core boundary (Fig. 1a). Implosion simulations [29] showed that the plasma temperature is typically within 0.3-1 keV. For simplicity we employ a uniform initial temperature 1 keV both for electrons and ions. Two 5 ps laser pulses each with 2.83 PW and 14.1 kJ are adopted, which counter-propagate from the two sides of the target along the ±x directions, respectively. This avoids too high energies of fast electrons generated by a single pulse of 5.66 PW. The two pulses generate two counter-propagating influxes of fast electrons to heat a single target core. The two pulses have a wavelength of 0.35 μm, linear polarization along the y direction, and intensity profile I=$I_0$ exp(-2r$^2$/w$^2$)f(ξ)sin$^2$(2πξ), where $I_0$=6$\times$10$^{20}$ W cm$^{-2}$, ξ=t ± x/c, w=17.3 μm, and the temporal profile f(ξ) is a plateau with 0.02 ps rising and decreasing edges. A 3.5 kT static magnetic field is imposed across the target along the x direction (see Methods for embedment of the magnetic field). We first perform a two-dimensional (2D) integrated PIC simulation with the above parameters, requiring 2 million core-hours on the supercomputer JUQUEEN. Three-dimensional (3D) integrated simulations with reduced scales were also performed to benchmark the 2D result, as shown below. The 2D simulation box size 224μm$\times$48μm in x$\times$y directions is taken and 49 electrons and ions per cell are employed to control the noise. The spatial resolution is 0.02 μm. Absorbing boundaries are adopted in all directions.

Figure 1c shows two counter-propagating strong influxes of fast electrons generated by the two laser pulses, which are well confined along the x direction by the external magnetic field. The influxes become wider with time because collisions cause transverse spreading and electrons with larger divergences are generated at higher-density plasma due to laser hole boring [24,25,30]. This suggests that one should adopt a laser spot size smaller than or comparable with the target core size for efficient heating, as we have taken. Later, the two influxes counteract around the x axis after they travel through the core center.

The two collimated influxes efficiently heat the core from its initial temperature of 1 keV to above 5 keV at 5.2 ps (Fig. 1b). The evolution of the average electron and ion temperatures in the core is plotted in Fig. 2c. At 5.2 ps the average temperatures are 5.13 keV and 5.04 keV for the electrons and ions, respectively. Inside the core the temperature distributions of ions shown in Fig. 1b are always similar to those of electrons due to the strong collisions. Outside of the core, their patterns are different. The ion temperatures display a double-lobe pattern, which shifts towards the high density core region with time. Besides, the spatial boundary of the core can be seen in Fig. 1b because a sudden change in density is taken, i.e., the value 400 g cm$^{-3}$ inside the core decreases to 133.3 g cm$^{-3}$ outside. More details of the dynamic evolution of the target temperatures and fast-electron currents are given in Supplementary Movie 1.

Figure 2a,b shows the energy coupling efficiencies from the heating laser pulses to the fast electrons and to the core. The fast electrons first gain 41% laser energy and then deposit 14.8% laser energy (corresponding to 4.2 kJ) to the core up to 5.2 ps. About 13% laser energy

(corresponding to 32% fast-electron energy) is lost due to the escape of the fast electrons with too high energy from the simulation region. Less than 4% laser energy is lost by the fast electrons escaping transversely, suggesting that the external magnetic field well confines the electron motion. As shown in Fig. 2c, the average energy of all fast electrons can be as high as 7 MeV at the time of 1 ps, which finally reduces to 2.3 MeV at 5.2 ps. The average energy is calculated at a given time with all the fast electrons arrived at ± 62 μm. This indicates that it is necessary to compute the electron generation in real time for the sequent calculation of the electron transport, as we have done here. Two factors are responsible for the energy reduction with time. First, the electrons with lower energy take longer time to arrive; Second, the energy of the fast electrons generated by the pulses goes down because the plasma density in front of the pulses grows with the laser hole boring [30] more deeply. Note that this reduction of fast electron energy does not lead to a reduction in the energy coupling efficiency from the pulses to the fast electrons since more fast electrons are produced in higher density plasma at later time. The reduced electron energy does not result in a continuous growth in the energy coupling from the pulses to the core. The first reason is that the collisions become weak as the core temperature rises. The second one is that the divergence of fast electrons generated in higher density plasma tends to increase, which decreases the probability to hit on the core, particularly for the electrons with lower energy.

**Optimization of imposed magnetic field strength**

The magnetic field strength of 3.5 kT used above has been optimized by a series of simulations with a single laser pulse, as depicted in Table 1. Note that the energy coupling from the laser to the fast electrons and the target core is almost the same between the cases with two counter-propagating laser pulses and with a single laser pulse, provided the laser parameters are the same. For 0.94—2.83 PW pulses with the spot radius of 17.3 µm, the optimized magnetic field strength is around 3—3.5 kT. Too weak magnetic fields cannot well confine the motion of divergent fast electrons. On the other hand, with increase of the magnetic field, the plasma density around the x axis or in the laser interaction area tends to be reduced because the background plasma electrons are expelled outwards by the collimated influx of the fast electrons. Once the plasma electron density is reduced, the laser hole boring effect will become stronger and the generated fast electrons will have higher energies, as shown in Table 1 for different magnetic strengths. With the higher-energy electrons, the energy coupling from the electrons to the core and also from the laser to the core tends to reduce, as observed in Table 1. Consequently, due to the two effects of the plasma density reduction and the magnetic confinement of the fast electrons, there is an optimized magnetic field strength.

**Parameter window for fusion ignition**

It is interesting to show a parameter window for fusion ignition for future experimental test. For this purpose, one needs to carry out full scale integrated simulations with a variety of parameters, which would incur huge computation expense. Here, we take an approximate way as follows. One can observe in Fig. 2b that the laser-to-core coupling slightly varies after 2.2 ps. Reasonably, one can use the coupling value at 2.2 ps (given in Table 1) to evaluate the laser energies required for ignition. Figure 3a displays the laser energies required to heat the core to 5 keV under different laser powers and magnetic field strengths. The corresponding laser durations required are presented in Fig. 3b. Notice that the laser durations should be shorter than 10 ps [6,12] to realize ignition. This condition can be met only when the power of each laser pulses is 1.88 PW or 2.83 PW according to Fig. 3b. With a lower power, e.g., 0.94 PW, the required laser durations are too long although the laser-to-core coupling is higher. One can also see that the laser energies and durations required for ignition weakly depend upon the magnetic field strength around the optimized value. This suggests that this magnetically assisted scheme is robust.

The results above are obtained with the laser pulses at $3\omega$ or 0.35 μm. Our additional simulations show that it is hardly possible to find a parameter window for achieving ignition conditions when the laser pulses are at 1.05 μm. This is because the generated fast electrons have too high energies [24,26] and consequently the laser-to-core coupling becomes too low. When the laser pulses are $2\omega$, ignition is still possible with increased laser pulse durations as compared with the $3\omega$ pulses. For example, if the laser power of 1.88 PW and the magnetic field of 3 kT are adopted, the durations of the $2\omega$ pulses should be 8.4 ps with the laser-to-core coupling of 12% obtained from our simulation; If the laser power of 2.83 PW and the magnetic field of 3.5 kT are adopted, the pulse durations should be 6.7 ps with the laser-to-core coupling of 10%. Note that we have chosen optimized target parameters with which a minimum laser energy is required for ignition and self-heating of the fusion core according to Ref. [23].

**Comparison between 2D and 3D simulations**

We have taken 2D integrated simulations above to save computation expense, since 3D ones will be orders of magnitude more expensive, which is almost impossible to implement currently.

Therefore, we reduce the target core radius to 4 μm and the density scalelength to 1.5 μm, and take a planar laser pulse. With the reduced scales we run both 3D and 2D integrated simulations for comparison. The 3D simulation box size is 25μm×10μm×10μm (x×y×z) with the spatial resolution 0.031 μm in all three directions. Eight electrons and ions are taken per cell. The external magnetic field of 3 kT is imposed along the x direction and the laser intensity is $I_0=10^{20}$ W cm$^{-2}$. The same parameters are taken in the corresponding 2D simulation except the particle number of 25 per cell. The simulation results are displayed in Fig. 4a,b. Both $\eta_{fast}$ and $\eta_{core}$ obtained from the 3D simulation are close to those from the 2D simulation in the whole simulation time. The maximal difference of $\eta_{core}$ is 1% and the difference of $\eta_{fast}$ is slightly larger, which agrees with the energy distributions of the fast electrons when they arrive at the injection point (at x=10 μm and see Methods). The energy distributions obtained from 2D and 3D simulations are close, but a little more electrons with energy between 3 MeV and 10 MeV are seen in the 2D simulation. These electrons with too high energy cannot efficiently deposit their energy to the 4-μm-radius core and consequently affect $\eta_{core}$ slightly. This comparison suggests that 2D integrated simulation results could approach the 3D ones when the laser spot radius is sufficiently large.

In the following, we compare 2D and 3D simulations with the laser spot radius of 14.1 μm, which is the minimum value used in Table 1, and the same target parameter as in Fig. 1 and Table 1. Due to the limit of computation resources, we evaluate the integrated simulation by two independent processes: the fast electron generation and collisional heating of fast electrons to plasma of 400 g cm$^{-3}$. First, we calculate the fast electron generation by one laser pulse using a conventional PIC simulation. The 3D simulation box size is 48μm×40μm×40μm (x×y×z) with the spatial resolution 0.031 μm in all three directions. The laser intensity and the magnetic field are taken as $I_0=4\times 10^{20}$ W cm$^{-2}$ and 3 kT, respectively. Figure 4c shows that $\eta_{fast}$ obtained from the 2D simulation is close to the 3D one in the whole simulation time and their maximal difference is 2%. This agrees with the result shown in Fig. 4d that the energy distributions of the fast electrons with energy below 3 MeV are similar between the 2D and 3D results. The main difference in the distributions appears at the high energy range above 5 MeV, which will slightly contribute to the core heating.

We then compare the collisional heating in 2D and 3D simulations, which is the dominant factor in the core heating by the fast electrons [21]. We take a uniform plasma with the density of 400 g cm$^{-3}$ and a size of 20 μm in three directions. An electron beam with a size 4μm in three directions is initially located at the front of the plasma. The beam electrons has the same velocity along the +x direction initially. We take periodic boundaries in three directions. Then the electrons will never escape the simulation box and always collide with the plasma continuously until reaching thermalization. In the simulations, the fields are suppressed and no laser pulse is incident. We change the initial energy $\varepsilon_0$ of the beam among 1 MeV, 5 MeV, and 10 MeV and obtain the evolution of the beam energy shown in Fig. 4e. One can see that the 2D results are in good agreement with the 3D ones, which is reasonable since the collision calculation [31] includes the full momentum space in both 2D and 3D simulation.

The comparisons between 2D and 3D simulations shown in Fig. 4 suggest that our 2D results in Figs. 1-3 and Table 1 should be reproducible in 3D simulations.

**Discussion**

In the present work, we show how the magnetically assisted scheme can realize ignition from full-scale PIC simulation, which self-consistently includes PW laser propagation, fast-electron generation, fast-electron transport to extreme dense regions, and target heating up to the ignition temperature for the first time. This work reveals an operating parameter range for the involved laser pulses and magnetic fields for experiments, which suggests that ignition can be realized with much reduced heating laser energy of 28 kJ. We also show that the 2ω or 3ω heating laser pulses should be taken to realize ignition.

We have taken a uniform magnetic field in our simulation. If the field is generated through the target compression process with a seed field imposed [19,20] (see Methods), the field should be non-uniform with the highest value on the target center. This may cause a magnetic mirroring effect to reduce the heating efficiency of the fast electrons, as shown in Refs. [32,33]. Strozzi *et al.* [32] proposed a hollow magnetic pipe scheme to overcome the mirroring effect. On the other hand, Johzaki *et al.* [33] found that the magnetic field with moderate non-uniformity can enhance the heating efficiency and proposed to optimize the timing between target compression and the seed magnetic field. Obviously, the optimization of the magnetic field structure will be an important topic to be considered in the future.

**Methods**

**Integrated PIC simulations.** We carry out 2D and 3D integrated PIC simulations with KLAPS code [27], where the integrated simulation approach, called two-system PIC scheme, for the whole heating process of fast ignition has been recently implemented [27]. It self-consistently includes electron generation by laser pulses as well as fast electron transport and energy deposition in a compressed target. This provides a straightforward way to obtain the energy coupling from the lasers to the target core.

In the two-system scheme, a conventional PIC system and a hybrid PIC system are used in a single simulation. Electron generation via laser plasma interaction is simulated by the conventional PIC system. When the fast electrons are transported to the regions (called the injection points) with the plasma density high enough and far away from the laser interaction zone (e.g., 250 $n_c$ in the simulation shown in Figs. 1 and 2 and 100 $n_c$ in other simulations), the data of these fast electrons are copied in real time to the hybrid PIC system with a reduced field solver as used in the hybrid PIC and the two-region PIC [34]. We define the fast electrons as these with energy above 0.3 MeV and longitudinal momentum $|p_x|>0.45\ m_e c$ (50keV) when they arrive at the injection points of x=± 62 μm in the simulation shown in Figs. 1 and 2. In other simulations the fast electrons are those moving forwardly with $p_x >0.45\ m_e c$ when a single laser pulse is taken. The hybrid PIC system calculates the subsequent transport of these fast electrons in the real target density conditions. In both systems macroparticles are taken to denote the background plasma and calculate Coulomb collisions [31]. Fourth order interpolation scheme [27] is adopted to reduce numerical noises.

The conventional PIC system has nearly no approximations in physical models used but it may appear numerical noise for high density plasma simulation. If the density is above 250 $n_c$ in the simulation shown in Fig. 1 and 2, it is lessened artificially to this value to avoid numerical noise [27]. The density reduction is suitable since the laser interaction zone is always far from such high density region, as seen in detailed discussion in Ref. [27]. The hybrid PIC model has an approximation in the field solver, i.e., the displacement current term in the Ampere's law is omitted. This approximation has been proved to be suitable to calculate electron transport in high density plasma, as found in the literature of hybrid PIC. In the revised hybrid PIC model used by us, which was proposed by Cohen et al. [34], the background plasmas are denoted by macroparticles rather than fluids in usually hybrid PIC and therefore the Coulomb collisions are computed by a Monte Carlo method with randomly-chosen particle pairs [31] to avoid using collision models between hot electrons and fluids.

**Embedment of the external magnetic field.** How the high magnetic field can be embedded into the compressed target is a key technological issue. There are two possible ways in our view. Firstly, one can follow the method used in magnetized ICF implosion [19,20]: a seed magnetic field of 8 T is imposed across the target and then it is magnified to about 8 kT through the target compression. The evolution of the magnetic field has the same timescale of implosion, which occurs on the nanosecond (ns) timescale far longer than our simulation period 5.2 ps. Besides, one can apply the directly-generated magnetic field of multi kT [18]. The magnetic field could be imposed at the same time when the ps, PW heating laser pulses are incident. The typical timing accuracy of kJ, ns laser systems is typically about 30-50 ps, which are applied to generate the magnetic fields [18]. The tens of ps timing to impose the magnetic field (with a ns life period) is far shorter than the timescale of the magnetic field evolution. Therefore, one can adopt either way to impose the magnetic field, where the evolution of the magnetic field topology can be neglected within the time scale of a few ps.

**Acknowledgements**

The authors are grateful to Prof. J. Meyer-ter-Vehn and Dr. B. Grant Logan for helpful comments and suggestions. The authors gratefully acknowledge the computing time granted by the JARA-HPC and VSR committees on the supercomputer JUQUEEN at Forschungszentrum Jülich. This work was supported by the National Basic Research Program of China (No. 2013CBA01500) and NSFC (Nos. 11375261, 11421064, 11375262 and 11135012).

**Author contributions**

W.M.W conceived the idea with P.G. and Z.M.S. W.M.W. carried out the PIC simulations. All authors contributed extensively to the data analysis. W.M.W, Z.M.S. and P.G. wrote the paper.

**Competing financial interests**

The authors declare no competing financial interests.

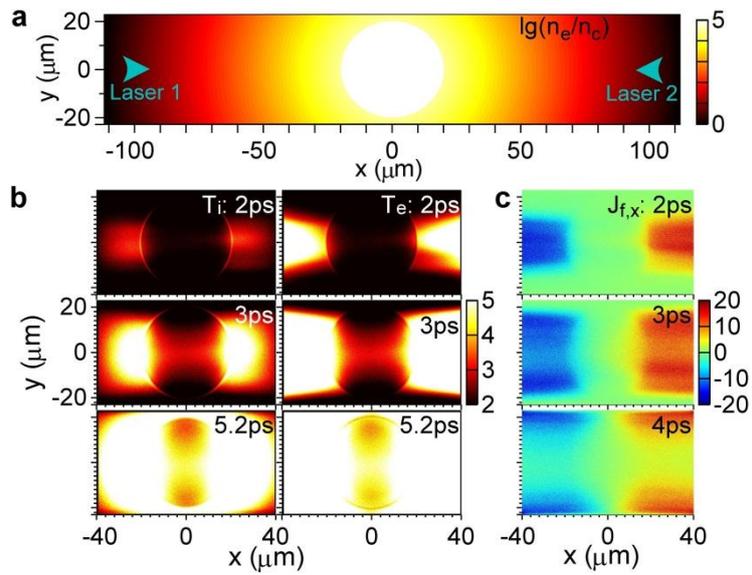

**Figure 1 | Target heating by 28.3 kJ counter-propagating laser pulses. a,** Distribution of the electron density lg($n_e/n_c$) of the pre-compressed target initially, where "Laser 1" and "Laser 2" denote two counter-propagating heating laser pulses. **b**, Snapshots of temperature distributions of ions ($T_i$/keV) and electrons ($T_e$/keV) at 2ps, 3ps, and 5.2ps, respectively. **c,** Fast-electron currents $J_{f,x}/ecn_c$ at 2ps, 3ps, and 4ps, respectively.

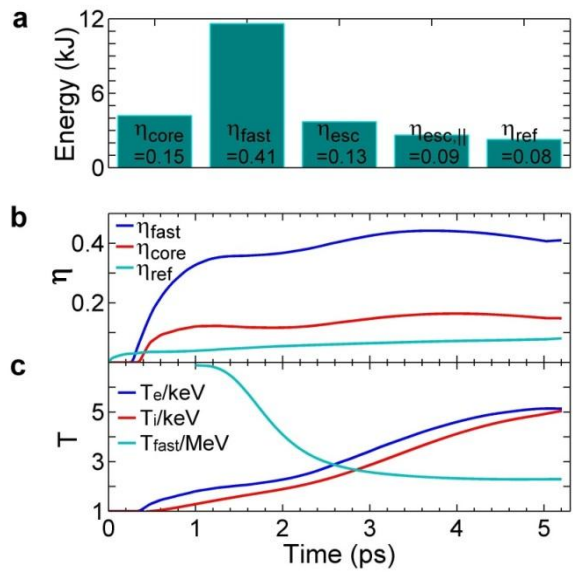

**Figure 2 | Energy coupling from the laser pulses to the fast electrons and the target core. a,** At 5.2 ps energies coupled to the core and the fast electrons from the laser pulses, energy of escaping fast electrons from simulation box, energy of longitudinally escaping electrons, and the reflected laser energy, where the corresponding efficiencies from the laser energy are also presented as $\eta_{core}$, $\eta_{fast}$, $\eta_{esc}$, $\eta_{esc,||}$, and $\eta_{ref}$, respectively. **b,** Temporal evolution of $\eta_{fast}$, $\eta_{core}$, and $\eta_{ref}$. **c,** Temporal evolution of temperatures (average energies) of the laser-generated fast electrons and the electrons and ions in the core.

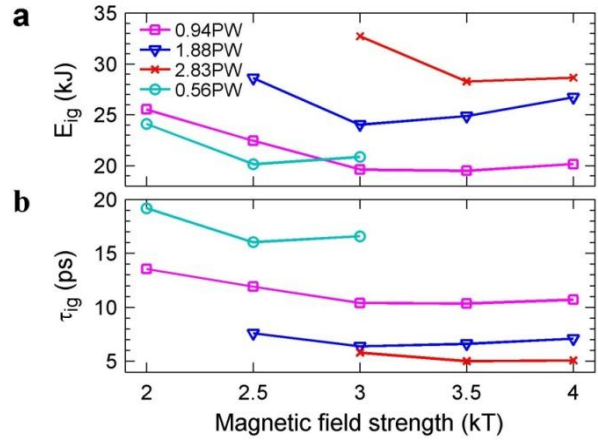

**Figure 3 | Ignition energy and time. a,** Total energy $E_{ig}$ of the two counter-propagating laser pulses required for achieving fusion ignition conditions as a function of the external magnetic strength and **b,** the corresponding laser durations required. Here, the power of each laser pulse is given with a spot radius of 17.3 µm, except the 0.56 PW one with a radius of 14.1 µm.

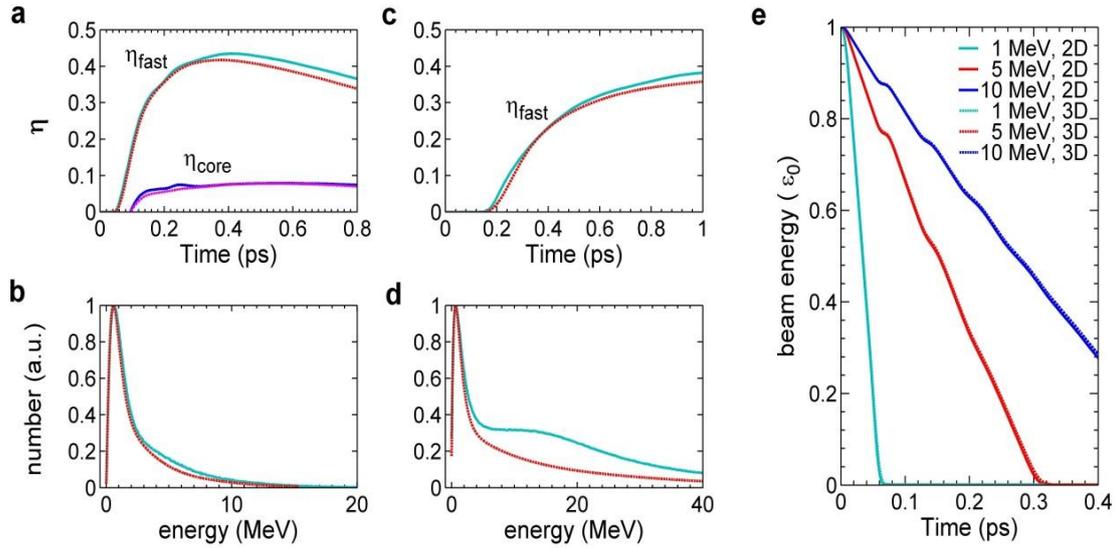

**Figure 4 | Comparison between 2D and 3D simulations. a,c,** Temporal evolution of the laser-to-fast-electron coupling $\eta_{fast}$ and laser-to-core coupling $\eta_{core}$ and **b,d,** the energy distributions of the fast electrons when they arrive at x=10 μm and x=36 μm, respectively, where the solid and broken lines correspond to 2D and 3D simulations, respectively. **a,b,** Integrated simulations with a planar laser pulse and **c,d,** simulations of fast electron generation with a laser pulse of 14.1 μm spot radius. **e,** Energy evolution of electron beams with different initial energies $\varepsilon_0$ due to Coulomb collision with plasma of 400 g cm$^{-3}$, where the solid and broken lines correspond to 2D and 3D simulations, respectively.

**Table 1 | Laser-to-core coupling efficiency and average energy of fast electrons.** They are given under different magnetic field strengths and laser powers, corresponding to laser intensities of $2\times10^{20}$ Wcm$^{-2}$, $4\times10^{20}$ Wcm$^{-2}$, $6\times10^{20}$ Wcm$^{-2}$, and $2\times10^{20}$ Wcm$^{-2}$, respectively. In the first three rows, the laser spot radius is fixed at 17.3 μm. In the last row, the spot radius is 14.1 μm. These values are obtained at simulation time of 2.2 ps.

|         | 2kT          | 2.5kT        | 3kT          | 3.5kT        | 4kT          |
|---------|--------------|--------------|--------------|--------------|--------------|
| 0.94 PW | 14.9%, 1.85MeV | 17.0%, 1.88MeV | 19.4%, 2.03MeV | 19.5%, 2.18MeV | 18.9%, 2.19MeV |
| 1.88 PW |              | 13.9%, 2.65MeV | 15.8%, 2.79MeV | 15.3%, 2.94MeV | 14.2%, 2.95MeV |
| 2.83 PW |              |              | 11.6%, 3.36MeV | 13.5%, 3.52MeV | 13.3%, 3.59MeV |
| 0.56 PW | 15.8%, 1.67MeV | 18.9%, 1.93MeV | 18.3%, 1.98MeV |              |              |

# Supplementary Information

## Fusion ignition via a magnetically-assisted fast ignition approach
W.-M. Wang, P. Gibbon, Z.-M. Sheng, Y. T. Li, and J. Zhang

**Dynamic evolution of fast-electron currents and target temperatures obtained from 2D integrated simulation**

(Please see **Movie S1** in the gif format.)

In order to show more details about the target core heating process, Movie S1 is presented, which is obtained from the same 2D integrated simulation under the conditions given in Fig. 1. The dynamic evolution of fast-electron currents and target temperatures is shown during 1 ps to 5.2 ps with a gap of 0.2 ps. Each frame of the movie includes the fast-electron current $J_{f,x}$ (top), electron temperature (middle), and ion temperature (bottom) at a time step.

**Two counter-propagating influxes of fast electrons obtained from 3D integrated simulation**

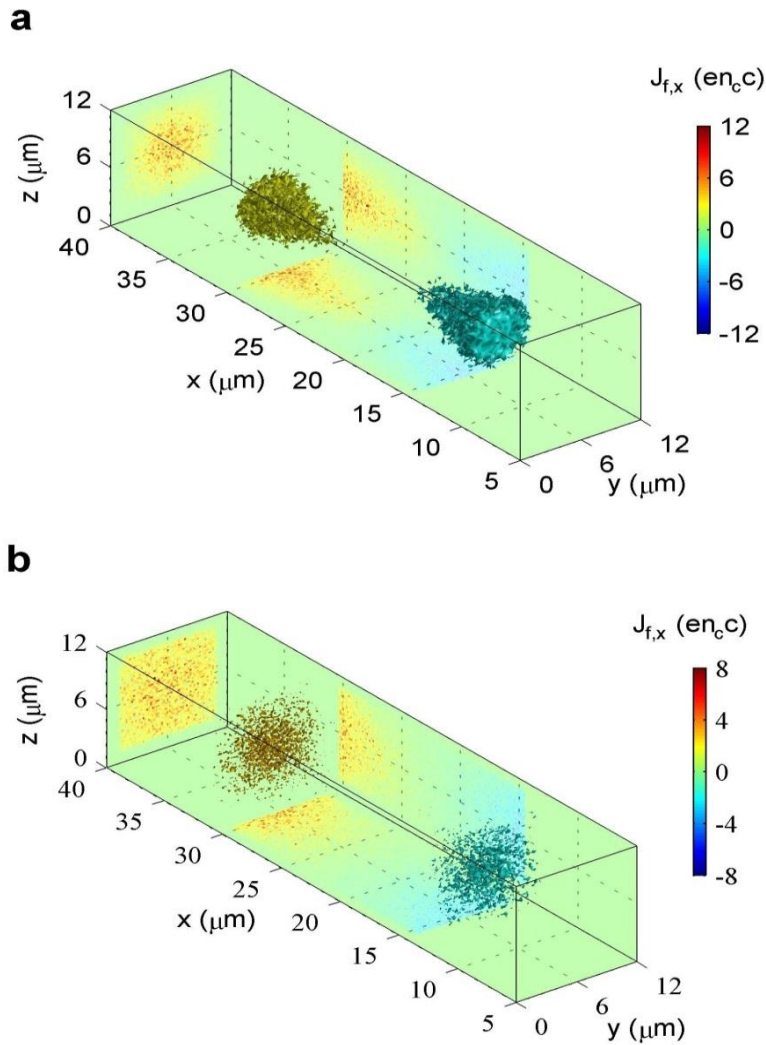

**Fig. S1:** Spatial distributions of fast-electron currents generated by two counter-propagating heating laser pulses at 0.2 ps, where (**a**) and (**b**) correspond to the cases with and without the external magnetic field of 3.5 kT, respectively.

    We run two 3D integrated simulations in a reduced scale to demonstrate the confinement of the external magnetic field to the fast electrons generated by two counter-propagating laser pulses with the peak intensity of $6\times10^{20}$ Wcm$^{-2}$. It is shown in Fig. S1a that the fast electrons are well confined to transport around the axis (y=z=0) when the magnetic field of 3.5 kT is imposed. Without the magnetic field, the fast electrons diffuse in the whole transverse space and the currents are almost uniform transversely, as observe in Fig. S1b. Note that the injection points are located at x=10 μm and x=30 μm from two sides and therefore, the fast-electron currents mainly distribute within x= 10 μm and x= 30 μm. We take the target with a uniform density 400 g cm$^{-3}$ within a 4-μm-radius core, and the surrounding density decreasing exponentially with a scalelength 1.8 μm. The laser spot radius is taken as 3 μm. The simulation box size is 40μm×12μm×12μm (x×y×z) with the spatial resolution 0.031 μm in all three directions. Eight electrons and ions per cell are taken. The simulation is done for the interaction time of just 0.2 ps.